\documentstyle[bezier]{article}
\begin{document}
\def \inbar{\vrule height1.5ex width.4pt depth0pt}
\def \xC{\relax\hbox{\kern.25em$\inbar\kern-.3em{\rm C}$}}
\def \xR{\relax{\rm I\kern-.18em R}}
\newcommand{\R}{\xR}
\newcommand{\C}{\xC}
\newcommand{\xZ}{Z \hspace{-.08in}Z}
\newcommand{\xbe}{\begin{equation}}
\newcommand{\be}{\begin{equation}}
\newcommand{\xee}{\end{equation}}
\newcommand{\ee}{\end{equation}}
\newcommand{\xbea}{\begin{eqnarray}}
\newcommand{\bea}{\begin{eqnarray}}
\newcommand{\xeea}{\end{eqnarray}}
\newcommand{\eea}{\end{eqnarray}}
\newcommand{\xnn}{\nonumber}
\newcommand{\nn}{\nonumber}
\newcommand{\xkt}{\rangle}
\newcommand{\kt}{\rangle}
\newcommand{\xbr}{\langle}
\newcommand{\br}{\langle}
\newcommand{\xcun}{\mbox{\footnotesize${\cal N}$}}
\newcommand{\cun}{\mbox{\footnotesize${\cal N}$}}
\newcommand{\cum}{\mbox{\footnotesize${\cal M}$}}
\title{Variational Sturmian Approximation: A nonperturbative
method of solving time-independent Schr\"odinger equation}
\author{Ali Mostafazadeh\thanks{E-mail address: 
amostafazadeh@ku.edu.tr}\\ \\
Department of Mathematics, Ko\c{c} University,\\
Rumelifeneri Yolu, 80910 Sariyer, Istanbul, TURKEY}
\date{ }
\maketitle

\begin{abstract}
A variationally improved Sturmian approximation for solving time-independent 
Schr\"odinger equation is developed. This approximation is used to obtain the 
energy levels of a quartic anharmonic oscillator, a quartic potential, and a 
Gaussian potential. The results are compared with those of the perturbation
theory, the WKB approximation, and the accurate numerical values. 
\end{abstract}
\vspace{2mm}



\section*{I.~Introduction}
Since the early days of quantum mechanics, the main technical tools for solving 
the time-independent Schr\"odinger equation have been the time-independent 
perturbation theory, the semiclassical or WKB approximation, and the 
variational method \cite{perturbation,landau}. Starting form the late 1950's, 
physical chemists and nuclear physicists have explored the use of what is 
called the {\em Sturmian basis functions} in solving this equation for a variety 
of potentials arising in molecular and atomic physics \cite{ro,sturmian}. 
Recently, Antonsen \cite{antonsen} and Szmytkowski and Zywicka-Mozeiko
\cite{poland} have studied the harmonic oscillator Sturmian functions. 
The purpose of the present article is to outline a general variationally improved 
Sturmian approximation scheme that provides a nonperturbative method of 
solving time-independent Schr\"odinger equation.

The organization of the article is as follows. In section~II, we give the definition 
of the Sturmian basis vectors, derive their general properties, and discuss the 
conventional Sturmian approximation. In section~III, we present an improved
Sturmian approximation which makes use of the variational method. In 
section~IV, we study the harmonic oscillator Sturmians and use them for the 
solution of time-independent Schr\"odinger equation in one dimension. In 
section~V, we apply our general results to some concrete problems, and
compare our results with those obtained using perturbation theory, the
WKB approximation, and the highly accurate numerical investigations. In 
particular, we obtain the energy levels of a quartic anharmonic oscillator, a 
quartic potential, and a Gaussian potential. Finally, in section~VI, we 
summarize our results and present our conclusions.

\section*{II.~Conventional Sturmian Approximation}

Consider the time-independent Schr\"odinger equation:
	\be
	H|E,a\kt=E|E,a\kt\;,
	\label{xe1}
	\ee
where $H$ is a self-adjoint Hamiltonian operator and $a$ is a degeneracy label. 

The method of Sturmian approximation is based on an expansion of the 
eigenvectors $|E,a\kt$ in terms of solutions $|\phi_\nu,\alpha\kt$ of the 
equation
	\be	
	(H_0+\beta_\nu V_0)|\phi_\nu,\alpha\kt={\cal E}|\phi_\nu,\alpha\kt\,,
	\label{xe2}
	\ee
where $H_0$ and $V_0$ are self-adjoint operators, $\beta_\nu$ 
and ${\cal E}$ are real scalar parameters, and $\alpha$ is a degeneracy label. 
Note that in order to obtain $|\phi_\nu,\alpha\kt$, one must fix ${\cal E}$ 
and solve Eq.~(\ref{xe2}) for $|\phi_\nu,\alpha\kt$. Clearly, every solution
$|\phi_\nu,\alpha\kt$ would correspond to a choice for the value of the `coupling constant' $\beta_\nu$. 

Suppose that $H_0={\vec p}^2/(2m)$ is the Hamiltonian for a free particle 
moving in  the configuration space $\R^d$, $V_0=V_0(\vec x)$ is a real 
interaction potential, and $H$ is a standard Hamiltonian of the form
	\be
	H=\frac{{\vec p}^2}{2m}+V(\vec x)\;,
	\label{e3}
	\ee
where $\vec p$ and $\vec x$ are momentum and position operators, respectively.
If $V_0(\vec x)$ tends to infinity as $|\vec x|\to\infty$, all the eigenstates of 
$V_0$ are bound states \cite{messiyah}. In this case, only for a discrete set of 
positive values of $\beta_\nu$, can we find square-integrable solutions 
$|\phi_\nu,\alpha\kt$ of Eq.~(\ref{xe2}). In this case, the label $\nu$ will take 
values in a discrete set which we shall choose to be $\{0,1,2,\cdots\}$.

The only difference between Eq.~(\ref{xe2}) and the eigenvalue equation for 
the potential $\beta_\nu V_0(\vec x)$ is that in the former  ${\cal E}$ is a fixed 
parameter which can be arbitrarily chosen. Therefore, a solution of 
Eq.~(\ref{xe2}) corresponds to a pair $(\beta_\nu,|\phi_{\nu},\alpha\kt)$. 

The vectors $|\phi_\nu,\alpha\kt$ are called the {\em Sturmian basis vectors} 
or simply the {\em Sturmians} \cite{ro}. They satisfy certain orthonormality 
conditions which we shall derive below. We should, however, note that the 
square-integrable Sturmians do not generally constitute a complete set of basis 
vectors of the Hilbert space \cite{poland}. There are certain potentials $V_0$,
such as the Coulomb potential, that lead to a complete set of square-integrable 
Sturmians \cite{weniger}.

Let us first note that the defining equation~(\ref{xe2}) does not determine 
$|\phi_\nu,\alpha\kt$ uniquely. This is reflected in the presence of the 
degeneracy label $\alpha$. What is uniquely determined by Eq.~(\ref{xe2}) is 
the degeneracy subspace ${\cal H}_\nu$ spanned by $\{|\phi_\nu,1\kt,
|\phi_\nu,2\kt,\cdots,|\phi_\nu,\ell_\nu\kt\}$, where $\ell_\nu$ is the degree of 
degeneracy, i.e., the number of linearly independent solutions of
Eq.~(\ref{xe2}) associated with a given (admissible) value of $\beta_\nu$. 
Clearly, we can construct an orthonormal basis of ${\cal H}_\nu$ and choose 
the Sturmian vectors $|\phi_\nu,\alpha\kt$ to be the basis vectors. In other 
words, without loss of generality, we can choose to work with the Sturmians 
$|\phi_\nu,\alpha\kt$ satisfying
	\be
	\br\phi_\nu,\alpha|\phi_\nu,\gamma\kt=\delta_{\alpha\gamma}\;,
	\label{orthono}
	\ee
where $\delta_{\alpha\gamma}$ denotes the Kronecker delta function. Clearly, 
any unitary transformation of ${\cal H}_\nu$ would lead to a new set of 
Sturmians satisfying (\ref{orthono}). Therefore, the condition (\ref{orthono}) 
reduces the freedom in the choice of the Sturmians $|\phi_\nu,\alpha\kt$, but 
does not eliminate it. 

Next, we evaluate the Hermitian adjoint of both sides of Eq.~(\ref{xe2}), 
change $(\nu,\alpha)$ to $(\mu,\gamma)$, and take the inner product of both 
sides of the resulting equation with $|\phi_\nu,\alpha\kt$. This yields
	\be
	\beta_\mu\br\phi_\mu,\gamma|V_0|\phi_\nu,\alpha\kt
	=\br\phi_\mu,\gamma|({\cal E}-H_0)|\phi_\nu,\alpha\kt\;.
	\label{e1}
	\ee
We can compute the right-hand side of this equation using Eq.~(\ref{xe2}). 
Substituting the result in (\ref{e1}), we find
	\be
	(\beta_\mu-\beta_\nu)\br\phi_\mu,\gamma|V_0|\phi_\nu,\alpha\kt=0\;.
	\label{e2}
	\ee
If we define
	\be
	N_{\nu}^{\gamma\alpha}:=\br\phi_\nu,\gamma|V_0|\phi_\nu,\alpha\kt\;,
	\label{N}
	\ee
then we can write Eq.~(\ref{e2}) in the form
	\be
	\br\phi_\mu,\gamma|V_0|\phi_\nu,\alpha\kt=
	N_\nu^{\gamma\alpha}\delta_{\mu\nu}\;.
	\label{ortho}
	\ee
Eq.~(\ref{ortho}) is the desired orthogonality property of the Sturmians. We 
can further simplify Eq.~(\ref{ortho}), by noting that the 
$\ell_\nu\times\ell_\nu$ matrix $N_{\nu}$ formed by 
$N_{\nu}^{\gamma\alpha}$ is a Hermitian matrix. This means that we can 
choose $|\phi_\nu,\alpha\kt$ in such a way that $N_{\nu}$ is a diagonal matrix. 
Making this choice, we have
	\bea	
	N_{\nu}^{\alpha\gamma}&=&
	N_{\nu}^{\alpha}\delta_{\alpha\gamma}\;,
	\label{e3new}\\
	\br\phi_\mu,\gamma|V_0(\vec x)|\phi_\nu,\alpha\kt&=&
	N_\nu^{\alpha}\delta_{\mu\nu}\delta_{\alpha\gamma}\;,
	\label{e4}
	\eea
where $N^\nu_\alpha$, with $\alpha\in\{1,2,\cdots,\ell_\nu\}$, are eigenvalues 
of the matrix $N_\nu$. Since $N_\nu$ is Hermitian, $N_\nu^\alpha$ are real.

In summary, we can choose a set of Sturmian vectors $|\phi_\nu,\alpha\kt$ 
which are eigenvectors of the matrices $N_\nu$. Therefore, for each value 
of $\nu$, $\{|\phi_\nu,1\kt,\cdots,|\phi_\nu,\ell_\nu\kt\}$ forms an orthonormal 
eigenbasis of $N_\nu$ in the degeneracy subspace ${\cal H}_\nu$. However, 
$|\phi_\nu,\alpha\kt$ with different values of $\nu$ are not orthogonal. Instead,
they satisfy a modified orthogonality condition, namely (\ref{e4}).

Now, let us expand a solution $|E,a\kt$ of the Schr\"odinger equation
(\ref{xe1}), in a Sturmian basis corresponding to a `solvable' potential 
$V_0$, i.e., seek solutions of the form
	\be
	|E,a\kt=\sum_{\nu=0}^\infty\sum_{\alpha=1}^{\ell_\nu}
	C_\nu^\alpha|\phi_\nu,\alpha\kt\;,
	\label{e6}
	\ee
where $C_\nu^\alpha$ are complex coefficients and $\nu$ is supposed to take
discrete values $0,1,2,\cdots$. Note that if the Sturmians $|\phi_\nu,\alpha\kt$
do not form a complete basis, then Eq.~(\ref{e6}) yields the eigenvectors that
belong to the span of $|\phi_\nu,\alpha\kt$. 

The {\em Sturmian Approximation of order
$N$} is the approximation in which one neglects all the coefficients 
$C_\nu^\alpha$ in Eq.~(\ref{e6}) but those with  $\nu$ belonging to a subset 
${\cal S}_{N+1}$ of nonnegative integers of order $N+1$. Alternatively, in 
considering the Sturmian approximation of order $N$, one confines the range 
of the indices (of type) $\nu$ to a fixed finite set ${\cal S}_{N+1}$. In this 
way, the infinite sum $\sum_{\nu=0}^\infty\cdots$ in Eq.~(\ref{e6}) is 
replaced by the finite sum $\sum_{\nu\in{\cal S}_{N+1}}\cdots$. We shall 
abbreviate the latter by $\sum_\nu$. The set ${\cal S}_{N+1}$ may, in 
principle, be chosen arbitrarily. We will comment on this choice in section~III.

Substituting (\ref{e6}) in the Schr\"odinger equation (\ref{xe1}) and making 
use of Eqs.~(\ref{xe2}) and (\ref{e3}), we find
	\be
	\sum_{\nu}\sum_{\alpha=1}^{\ell_\nu} C_\nu^\alpha
	\left(E-{\cal E}-V+\beta_\nu V_0\right)|\phi_\nu,\alpha\kt=0.
	\label{e8}
	\ee
Now, evaluating the inner product of both sides of this equation with 
$|\phi_\mu,\gamma\kt$ and using the orthogonality relation~(\ref{e4}), 
we obtain
	\be
	\sum_{\nu}\sum_{\alpha=1}^{\ell_\nu}\left[(E-{\cal E})
	T^{\gamma\alpha}_{\mu\nu}-(W^{\gamma\alpha}_{\mu\nu}-
	\beta_\nu N^\alpha_\nu\delta_{\mu\nu}\delta_{\gamma\alpha})\right]
	C^\alpha_\nu=0\;.
	\label{e9}
	\ee
Here we have introduced
	\bea
	T_{\mu\nu}^{\gamma\alpha}&:=&
	\br\phi_\mu,\gamma|\phi_\nu,\alpha\kt\;,
	\label{T}\\
	W_{\mu\nu}^{\gamma\alpha}&:=&
	\br\phi_\mu,\gamma|V|\phi_\nu,\alpha\kt\;.
	\label{W}
	\eea
We can express Eq.~(\ref{e9}) in a more compact form, if we use a single 
label for the pair $(\nu,\alpha)$. Introducing $\cun:=(\nu,\alpha)$ and 
$\cum:=(\mu,\gamma)$, we write  Eq.~(\ref{e9}) in the form
	\be
	\sum_{\cun} \left[(E-{\cal E})T_{\cum\cun}-S_{\cum\cun}\right]
	C_{\cun}=0\;,
	\label{e10}
	\ee
where
	\bea
	T_{\cum\cun}&=&T_{\mu\nu}^{\gamma\alpha}\;,~~~~
	S_{\cum\cun}=W_{\cum\cun}-\beta_\nu N_{\cun}\delta_{\cum\cun}\;,
	\label{e11.1}\\
	W_{\cum\cun}&=&W_{\mu\nu}^{\gamma\alpha}\;,~~~~
	N_{\cun}=N^\alpha_\nu\;.
	\label{e11.2}
	\eea

Eqs.~(\ref{e10}) form a linear system of homogeneous first order algebraic 
equations for $C_{\cun}$. This system has a nontrivial solution provided that 
the determinant of the matrix of coefficients vanishes, i.e.,
	\be
	\det\left[(E-{\cal E})T-S\right]=0\;.
	\label{e12}
	\ee
Here $T$ and $S$ are matrices with entries $T_{\cum\cun}$ and 
$S_{\cum\cun}$, respectively.

Solving Eq.~(\ref{e12}), we can express $E$ in terms of ${\cal E}$, 
$\beta_\nu$, and the Sturmians $|\phi_\nu,\alpha\kt$. Now, we recall that for a 
fixed choice of  $V_0$, the coupling constants $\beta_\nu$ and the 
corresponding Sturmians $|\phi_\nu,\alpha\kt$ depend on the parameter 
${\cal E}$. Therefore, Eq.~(\ref{e12}) yields $E$  as a function of 
${\cal E}$. Furthermore, substituting the value of $E=E({\cal E})$ 
obtained by solving (\ref{e12}) in (\ref{e10}) and solving for the coefficients
$C_{\cun}$, we obtain an expression for the eigenvector $|E,a\kt$ that
involves ${\cal E}$. As we discuss in the following section, the fact that the 
Sturmian approximation yields the eigenvalues and the eigenvectors of the 
Hamiltonian as functions of a free parameter seems to have been overlooked. 
This is mainly because there is a choice for the parameter ${\cal E}$ that 
simplifies the calculations. 

It  should also be emphasized that Eq.~(\ref{e12}) is an algebraic equation of 
order $N+1$. Therefore, in general, it has $N+1$ solutions. This can be 
understood by noting that in the Sturmian approximation one actually
approximates the Hilbert space by a finite-dimensional vector space. 
Consequently, the Hamiltonian is replaced with a matrix with a finite number 
of eigenvalues.

\section*{III.~Variational Sturmian Approximation}

In general, the accuracy of the Sturmian approximation depends on the 
following factors.
	\begin{itemize}
	\item[] {\bf 1. Choice of $V_0$:} In practice, $V_0$ must be one of the 
exactly solvable potentials. Therefore, the available choices for $V_0$ are few 
in number. For the potentials $V$ with bound states, we can choose $V_0$
to be a harmonic oscillator potential. For example, for the quartic anharmonic
oscillator 
	\be
	V(x)= \frac{k}{2}\,x^2 +\epsilon\, x^4\;,
	\label{quartic-v}
	\ee
we shall take
	\be
	V_0(x)=\frac{k}{2}\,x^2\;.
	\label{quartic-v0} 
	\ee
Similarly, for the Gaussian potential
	\be
	V(x)=-\lambda e^{-\epsilon x^2/2}\;,
	\label{gaussian}
	\ee
we shall take 
	\be
	V_0(x)=\frac{1}{2}\lambda\epsilon x^2-\lambda\;.
	\label{sho}
	\ee
This will enable us to compare the results of the Sturmian approximation
with those of the perturbation theory, for in the limit $\epsilon\to 0$, 
$V(r)\to V_0(r)$. Note that multiplying $V_0$ by a positive real number does
not change the results of the Sturmian approximation. This is simply because
we can always absorb such a number in the definition of $\beta_\nu$. 
	\item[] {\bf 2. Choice of the Sturmians included in the sum (\ref{e10}):} 
This is also directly related to the choice of the potential $V_0$. If $V_0$ is 
obtained from $V$ by a limiting process as in the case of the 
potentials~(\ref{quartic-v}) and (\ref{gaussian}), then a natural choice for the 
computation of the $n$-th energy eigenvalue $E_n$ and the corresponding 
eigenvectors $|E_n,\alpha\kt$ is to include the $|\phi_\nu,\alpha\kt$ with $\nu$
equal or close to $n$. In particular, in the Sturmian approximation of order zero, we have
	\be
	|E_n,a\kt=
	\sum_{\alpha=1}^{\ell_n}C^\alpha_n|\phi_n,\alpha\kt\;.
	\label{e10-new}
	\ee
	\item[] {\bf 3. Choice of the parameter ${\cal E}$:}
The conventional choice \cite{sturmian,antonsen} for the parameter 
${\cal E}$ is ${\cal E}=E$. This simplifies Eq.~(\ref{e12}) considerably. The 
basic idea pursued in this article is the fact that this simplification does not 
necessarily justify the conventional choice for ${\cal E}$.
	\end{itemize}

It is well-known \cite{landau} that the eigenvalue equation (\ref{xe1})
is equivalent to the variational equation
	\be
	\frac{\delta}{\delta\br\psi|}\left(\frac{\br\psi|H|\psi\kt}{\br\psi|\psi\kt}
	\right)=0\;.
	\label{e13}
	\ee
In other words, the eigenvalues $E$ are the minima of the expectation value 
	\[\br H\kt:=\frac{\br\psi|H|\psi\kt}{\br\psi|\psi\kt}\,,\]
and the eigenvectors are the vectors $|E\kt=|E,a\kt$ that minimize 
$\br H\kt$. This observation suggests that the most efficient choice for the 
parameter ${\cal E}$ appearing in the Sturmian approximation is the one that 
minimizes $E=E({\cal E})$. Therefore, the most reliable Sturmian 
approximation is obtained by choosing ${\cal E}$ to be a solution of
	\be
	\frac{dE}{d{\cal E}}=0\;.
	\label{e14}
	\ee
If this equation does not have a solution, then one must either make another 
choice for the set ${\cal S}_{N+1}$ or proceed with a higher order Sturmian 
approximation.

\section*{IV.~Variational Sturmian Approximation Using Harmonic Oscillator Sturmians}

Consider a quantum system with the configuration space $\R$, a standard 
Hamiltonian~(\ref{e3}), and a real-valued potential $V=V(x)$. Suppose that
the system has an infinite number of bound states with nondegenerate energy
eigenvalues $E_n$. Here $n\in\{0,1,2,\cdots\}$ and $E_0$ stands for the 
ground state.

Now, consider the Sturmian basis vectors associated with a harmonic oscillator
\cite{antonsen,poland},
	\be
	V_0=V_0(x)=\frac{k}{2}\,x^2\;.
	\label{f1}
	\ee
In order to solve Eq.~(\ref{xe2}) for this choice of $V_0$, we introduce
	\bea
	\omega_\nu&:=&\left(\frac{\beta_\nu k}{m}\right)^{1/2}\,,
	\label{f1.1}\\
	\alpha_\nu&:=&\frac{m\omega_\nu}{\hbar}\,,
	\label{alpha}\\
	a_\nu&:=&\left(\frac{\alpha_\nu}{2}\right)^{1/2}
	\left(x+\frac{ip}{\hbar\alpha_\nu}\right)\;.
	\label{f1.2}\\
	|\ell\kt_\nu&:=&\frac{1}{\sqrt \ell !}\,a^{\dagger\ell}_\nu|0\kt_\nu\;,
	\label{f1.3}
	\eea
where $|0\kt_\nu$ is the normalized real ground state vector for a harmonic 
oscillator with mass $m$ and frequency $\omega_\nu$. That is
	\be
	\br x|0\kt_\nu:=\left(\frac{\alpha_\nu}{\pi}\right)^{1/4}
	e^{-\alpha_\nu x^2/2}.
	\label{f1.4}
	\ee
In view of the similarity of Eq.~(\ref{xe2}) with the eigenvalue equation for
the potential $\beta_\nu V_0$, we can easily deduce that
	\bea
	{\cal E}&=&\hbar\omega_\nu(\nu+\frac{1}{2})\;,
	\label{f2.1}\\
	|\phi_\nu\kt&=&|\nu\kt_\nu
	\label{f2.2}
	\eea
where $\nu\in\{0,1,2,\cdots\}$. 

We can invert Eqs.~(\ref{f2.1}) and (\ref{f1.1}) to express $\omega_\nu$ and
$\beta_\nu$ in terms of ${\cal E}$. This yields
	\bea
	\omega_\nu&=&\frac{2{\cal E}}{\hbar(2\nu+1)}\;,
	\label{f2}\\
	\beta_\nu&=&	\frac{4m{\cal E}^2}{\hbar^2k(2\nu+1)^2}\;.
	\label{f3}
	\eea
Substituting Eq.~(\ref{f2})  in (\ref{alpha}), we have
	\be
	\alpha_\nu=\frac{2m{\cal E}}{\hbar^2(2\nu+1)}=
	\frac{\alpha_0}{2\nu+1}\;.
	\label{alpha=}
	\ee

Next, we compute the term 
$\beta_\nu N_\nu=\beta_\nu\br\phi_\nu|V_0|\phi_\nu\kt$.
We can use the properties of the annihilation operator $a_\nu$, namely
	\be
	a_\nu|\ell\kt_\nu=\sqrt{\ell}|\ell-1\kt_\nu\;,~~~
	a_\nu^\dagger|\ell\kt_\nu=\sqrt{\ell+1}|\ell+1\kt_\nu\;,~~~
	x=(2\alpha_\nu)^{-1/2}(a_\nu+a_\nu^\dagger),
	\label{f11.2}
	\ee
and the orthonormality of $|\ell\kt_\nu$ to compute
	\be	
	~_\nu\br\ell|x^2|\nu\kt_\nu= (2\alpha_\nu)^{-1}\left[
	(2\nu+1)\delta_{\ell,\nu}+\sqrt{(\nu+1)(\nu+2)}\,\delta_{\ell,\nu+2}+
	\sqrt{\nu(\nu-1)}\,\delta_{\ell,\nu-2}\right]\;.
	\label{f12}
	\ee
In view of  Eqs.~(\ref{f1}) and (\ref{f12}), we obtain, after some remarkable simplifications,
	\be
	\beta_\nu N_\nu=\frac{\cal E}{2}\;.
	\label{f6}
	\ee
\subsection*{Variational Sturmian Approximation of Order Zero}

For the variational Sturmian approximation of order zero, $\nu=n$, and 
Eq.~(\ref{e12}) takes the form
	\be
	(E_n-{\cal E})T-S=0\;,
	\label{f4}
	\ee
where
	\be
	T=\br\phi_n|\phi_n\kt=1\,, ~~~~ S=W-\beta_n N_n\,,
	~~~~ W=\br\phi_n|V|\phi_n\kt\;.
	\label{f5.2}
	\ee
According to Eqs.~(\ref{f4}), (\ref{f5.2}), and (\ref{f6}), the energy eigenvalues
$E_n$ are given by
	\be
	E_n=W+\frac{\cal E}{2}\;.
	\label{f7}
	\ee

Next, we fix the parameter ${\cal E}$ using Eq.~(\ref{e14}).
This requires the computation of $dW/d{\cal E}$. We first evaluate the 
variation of $W$,
	\bea
	\delta W&=&(\delta\br\phi_n|)V|\phi_n\kt+ \br\phi_n|V(\delta|\phi_n\kt)
	\nn\\
 	&=&2 \br\phi_n|V (\delta|\phi_n\kt)\nn\\
	&=&2\sum_{\ell=0}^\infty ~_n\br n|V|\ell\kt_n~_n\br\ell|(\delta|n\kt_n)\;.
	\label{f8}
	\eea
Here we have made use of Eq.~(\ref{f2.2}), the fact that the Sturmians and 
the potential $V$ are real and $|\ell\kt_n$ form a complete set of basis vectors.

We can compute $~_n\br\ell|(\delta|n\kt_n)$ using the eigenvalue equation
	\be
	(H_0+\beta_n V_0)|j\kt_n={\cal E}_j|j\kt_n\;,
	\label{f9}
	\ee
where ${\cal E}_j=\hbar\omega_n(j+1/2)$. Taking the variation of both sides
of this equation and computing the inner product with $|\ell\kt_n$, we find
	\be
	~_n\br \ell|\delta|j\kt_n =\frac{ ~_n\br\ell| V_0|j\kt_n \,(\delta\beta_n) }{
	{\cal E}_j-{\cal E}_\ell} ~~~~~~{\rm for}~~\ell\neq j\;.
	\label{f10}
	\ee
Furthermore, using the fact that the eigenfunctions $\br x|\ell\kt_n$ are real, we
can easily show that 
	\be
	~_n\br \ell|(\delta|\ell\kt_n)=0\;.
	\label{f11}
	\ee

Eqs.~(\ref{f8}), (\ref{f10}) and (\ref{f11}) reduce the computation of 
$\delta W$ to that of
	\be
	 ~_n\br\ell| V_0|n\kt_n =\frac{k}{2}  ~_n\br\ell|x^2|n\kt_n  \;.
	\label{f11.1}
	\ee
We have already computed $~_n\br\ell|x^2|n\kt_n$ in Eq.~(\ref{f12}). 
Substituting this equation in Eq.~(\ref{f11.1}) and using Eqs.~(\ref{f10})
and (\ref{f8}), we find, after some remarkable cancellations,
	\be
	\delta W= \left(\frac{\delta{\cal E}}{2{\cal E}}\right) \left[
	\sqrt{n(n-1)}~_n\br n|V|n-2\kt_n-\sqrt{(n+1)(n+2)}~_n\br n|V|n+2\kt_n
	\right]\;.
	\label{f13}
	\ee

Now, in view of Eqs.~(\ref{f7}) and (\ref{f13}),
	\[\frac{dE_n}{d{\cal E}}=\left(\frac{1}{2{\cal E}}\right)
	\left[	\sqrt{n(n-1)}~_n\br n|V|n-2\kt_n - 
	\sqrt{(n+1)(n+2)}~_n\br n|V|n+2\kt_n\right]+\frac{1}{2}\;.\]
Substituting this equation in Eq.~(\ref{e14}) yields
	\be
	{\cal E}=\sqrt{(n+1)(n+2)}~_n\br n|V|n+2\kt_n-
	\sqrt{n(n-1)}~_n\br n|V|n-2\kt_n\;.
	\label{f14}
	\ee
Note that the right-hand side of this equation also involves ${\cal E}$. This
is because $|\ell\kt_n$ depend on ${\cal E}$. 

Using Eqs.~(\ref{f14}) and (\ref{f7}) we can express the energy eigenvalue 
$E_n$ in terms of $V$. This yields
	\be
	E_n= ~_n\br n|V|n\kt_n +\frac{1}{2}\,
	\left [\sqrt{(n+1)(n+2)} ~_n\br n|V|n+2\kt_n -
	\sqrt{n(n-1)}~_n\br n|V|n-2\kt_n\right]\;.
	\label{f15}
	\ee
For the ground state $n=0$~ and Eq.~(\ref{f15}) reduces to
	\be
	E_0=~_0\br 0|V|0\kt_0+\frac{1}{\sqrt{2}}~_0\br 0|V|2\kt_0\;.
	\label{f16}
	\ee
Note that the vectors $|\ell\kt_n$ appearing in Eqs.~(\ref{f15}) and 
(\ref{f16}) are those of Eq.~(\ref{f1.3}) with ${\cal E}$ being a
solution of Eq.~(\ref{f14}).

Eqs.~(\ref{f15}) and (\ref{f16}) are of limited importance. In practice, one 
obtains the energy eigenvalue $E_n$ by substituting the solution of 
Eq.~(\ref{f14}) in Eq.~(\ref{f7}). 

The variational Sturmian approximation of order zero, as outlined above,
is a valid approximation scheme, if Eq.~(\ref{f14}) has a unique
positive solution~${\cal E}$. If such a solution does not exist, one may 
attempt to construct higher order variational Sturmian approximations.
As we shall see in section~V, for all the potentials that we have considered,
Eq.~(\ref{f14}) has a unique positive solution. This is very remarkable, for
this equation turns out to be an algebraic equation of order three for
the quartic anharmonic oscillator and the quartic potential, and of order four 
for the Gaussian potential. 

\subsection*{Variational Sturmian Approximation of Order One}

In the variational sturmian approximation of order one, the number of
Sturmians contributing to the eigenvector $|E\kt$ is two. We shall denote
them by $|\phi_n\kt$ and $|\phi_m\kt$.

The matrices $T$ and $W$ are Hermitian $2\times 2$ matrices. They can 
be written in the form
	\be
	T=\left(\begin{array}{cc}
	1&t^*\\
	t&1\end{array}\right)\,,~~~~
	W=\left(\begin{array}{cc}
	v_n&w^*\\
	w&v_m\end{array}\right)\,,
	\label{f18}
	\ee
where we have used the fact that the Sturmians are normalized and introduced
	\be
	t:= \br\phi_m|\phi_n\kt, ~~~v_\nu:=\br\phi_\nu|V|\phi_\nu\kt, ~~~
	w:=\br\phi_m|V|\phi_n\kt.
	\label{f19}
	\ee
Next, we construct the matrix $S$. In view of Eqs.~(\ref{e11.1}) and (\ref{f18}), 
	\be
	S=\left(\begin{array}{cc}
	v_n-\beta_n N_n&w^*\\
	w&v_m-\beta_m N_m\end{array}\right)\,.
	\label{f20}
	\ee

Note that because the Sturmians for the harmonic oscillator are real-valued, 
$t$ and $w$ are real-valued functions of the parameter ${\cal E}$. In 
particular, the matrices $T,~W$ and $S$ are real and symmetric.

Substituting Eqs.~(\ref{f18})  and (\ref{f20}) in the Eq.~(\ref{e12}), making 
use of Eq.~(\ref{f6}), and simplifying the resulting expression, we find
	\be
	A (E-\frac{\cal E}{2})^2+B(E-\frac{\cal E}{2})+C=0\;,
	\label{f21}
	\ee
where
	\be
	A:=1-t^2,~~~B:=t(t{\cal E}+2w)-(v_n+v_m),~~~C:=v_nv_m-(
	\frac{t{\cal E}}{2}+w)^2.
	\label{f22}
	\ee
Note that the coefficients $A$, $B$, and $C$ are functions of ${\cal E}$.
Eq.~(\ref{f21}) can be easily solved to express $E$ in terms of ${\cal E}$.
The result is
	\be
	E=E_\pm:=\frac{\cal E}{2}+\frac{-B\pm\sqrt{B^2-4AC}}{2A}\;.
	\label{f23}
	\ee

The next step is to determine ${\cal E}$ using the variational principle, i.e., 
setting $dE/d{\cal E}=0$. The resulting formulas are complicated and we shall
not include them here. 

We conclude this section with the following remarks.
	\begin{itemize}
	\item[1.] As seen from Eq.~(\ref{f23}), the first order Sturmian 
	approximation leads to a pair of energy eigenvalues. For the potentials
	which are related to $V_0$ via a limiting process, one expects these
	two eigenvalues to be those labelled by $m$ and $n$. That is, for $n<m$,
		\be
		E_n=E_-,~~~~E_m=E_+\;.
		\label{f24}
		\ee
	\item[2.] In the variational Sturmian approximation of order one, there are 
	two variational equations $dE_\pm/d{\cal E}=0$. It is not clear whether 
	these equations lead to a unique minimum for $E_\pm({\cal E})$ with a 
	positive value for ${\cal E}$. As we shall show in the following sections, 
	for all the specific examples that we have considered each of these 
	equations lead to a unique minimum with a positive value for ${\cal E}$. 
	Lack of such a solution may be interpreted as the failure of the variational 
	Sturmian 	approximation of order one.
	\end{itemize}

\subsection*{Variational Sturmian Approximation of Order Two}

In the variational sturmian approximation of order two, one uses three 
Sturmians to expand the energy eigenvectors $|E\kt$. We shall denote these by 
$|\phi_{n_\ell}\kt$ where $n_\ell\in{\cal S}_3:=\{n_1,n_2,n_3\}$.

The matrices $T$, $W$, and $S$ are given by
	\bea
	T&=&\left(\begin{array}{ccc}
	1 & t_1^* & t_2^*\\
	t_1 & 1 & t_3^* \\
	t_2 & t_3 & 1 \end{array}\right),~~~~
	W=\left(\begin{array}{ccc}
	v_3 & w_1^* & w_2^*\\
	w_1 & v_2 & w_3^* \\
	w_2 & w_3 & v_1 \end{array}\right)\;,
	\label{f25.1}\\
	S&=&\left(\begin{array}{ccc}
	v_3 -\beta_{n_3}N_{n_3} & w_1^* & w_2^*\\
	w_1 & v_2-\beta_{n_2}N_{n_2} & w_3^* \\
	w_2 & w_3 & v_1-\beta_{n_1}N_{n_1} \end{array}\right)\;,
	\label{f25.3}
	\eea
where we have used the fact that $|\phi_{n_\ell}\kt$ are normalized and introduced
	\bea
	t_1&:=&\br\phi_{n_2}|\phi_{n_3}\kt,~~~~
	t_2:=\br\phi_{n_1}|\phi_{n_3}\kt,~~~~
	t_3:=\br\phi_{n_1}|\phi_{n_2}\kt,~~~~
	v_{\ell}:=\br\phi_{n_\ell}|V|\phi_{n_\ell}\kt,
	\label{f26.1}\\
	w_1&:=&\br\phi_{n_2}|V|\phi_{n_3}\kt,~~~~
	w_2:=\br\phi_{n_1}|V|\phi_{n_3}\kt,~~~~
	w_3:=\br\phi_{n_1}|V|\phi_{n_2}\kt.
	\label{f26.3}
	\eea
Because the harmonic oscillator Sturmian functions are real-valued, $t_\ell$, 
$v_\ell$ and $w_\ell$ are real, and $T$, $W$, and $S$ are real symmetric 
matrices. 

In view of Eq.~(\ref{f6}), we can write the secular equation~(\ref{e12}) in
the form
	\be
	A (E-\frac{\cal E}{2})^3 +B(E-\frac{\cal E}{2})^2+
	C(E-\frac{\cal E}{2})+D=0\;,
	\label{f27}
	\ee
where
	\bea
	A&:=&1-\sum_{\ell=1}^3 t_\ell^2+2t_1t_2t_3\;,
	\label{f28.1}\\
	B&:=&\sum_{\ell=1}^3[(t_\ell^2-1)v_\ell+2t_\ell\xi_\ell]
	-2(t_1t_2\xi_3+t_3t_1\xi_2+t_2t_3\xi_1)\;,
	\label{f28.2}\\
	C&:=&v_1v_2+v_2v_3+v_3v_1+2(t_1\xi_2\xi_3+t_3\xi_1\xi_2+
	t_2\xi_3\xi_1)-\sum_{\ell=1}^3(\xi_\ell^2+2t_\ell v_\ell\xi_\ell)\;,
	\label{f28.3}\\
	D&:=&\sum_{\ell=1}^3v_\ell\xi^2_\ell-2\xi_1\xi_2\xi_3-v_1v_2V_3\;,
	\label{f28.4}\\
	\xi_\ell&:=&\frac{1}{2}\,t_\ell {\cal E}+w_\ell\;.
	\label{f29}
	\eea
Eq.~(\ref{f27}) has, in general, three solutions. The desired eigenvalues are the
minima of these solutions corresponding to positive values of ${\cal E}$. Again 
for the cases where $V$ is related to $V_0$ by a limiting process the minima of 
the solutions of Eq.~(\ref{f27}) correspond to $E_{n_1},E_{n_2}$, and
$E_{n_3}$.

\section*{V.~Applications}

In this section, we apply our general results to compute the energy eigenvalues
of a quartic anharmonic oscillator, a quartic potential, and a Gaussian potential.

\subsection*{The Quartic Anharmonic Oscillator}

Consider the potential
	\be
	V(x)=\frac{k}{2}\, x^2+\epsilon x^4\;.
	\label{g1}
	\ee
In order to obtain the energy levels of this potential using variational Sturmian 
approximation of order zero, we need to calculate $~_n\br n|V|\ell\kt_n$. We 
first use Eqs.~(\ref{f11.2}) to compute
	\bea
	~_n\br n|x^4|\ell\kt_n&=&(2\alpha_n)^{-2}\left[
	3(2n^2+2n+1)\delta_{\ell,n}+4(n+1)\sqrt{(n+1)(n+2)}\,
	\delta_{\ell,n+2}+\right.\nn\\
	&&2(2n-1)\sqrt{n(n-1)}\,\delta_{\ell,n-2}+
	\sqrt{ (n+1) (n+2) (n+3) (n+4)}\,\delta_{\ell,n+4}+\nn\\
	&&	\left.\sqrt{(n-3) (n-2) (n-1)n} \,\delta_{\ell,n-4}\right].
	\label{g2}
	\eea
 In view of this equation and Eqs.~(\ref{f12}) and (\ref{g1}), 
	\bea
	&&W=~_n\br n|V|n\kt_n = \frac{(2n+1)k }{4\alpha_n}+
	\frac{3(2n^2+2n+1)\epsilon}{4\alpha_n^2}\;,
	\label{g3.1}\\
	&&~_n\br n|V|n+2\kt_n =
	\sqrt{(n+1)(n+2)}\,\left(\frac{ k}{4\alpha_n}+
	\frac{(n+1)\epsilon}{\alpha_n^2}\right)\;,
	\label{g3.2}\\
	&&~_n\br n|V|n-2\kt_n =
	\sqrt{n(n-1)}\,\left(\frac{ k}{4\alpha_n}+
	\frac{(2n-1)\epsilon}{2\alpha_n^2}\right)\;,
	\label{g3.3}
	\eea

Next, we substitute Eqs.~(\ref{g3.2}) and (\ref{g3.3}) in Eq.~(\ref{f14}). 
Using Eq.~(\ref{alpha=}), we then obtain
	\be
	{\cal E}^3-p_n{\cal E}-q_n=0\;,
	\label{g4}
	\ee
where
	\be
	p_n:= \left(\frac{\hbar^2 k}{m}\right)(n+\frac{1}{2})^2,~~~~
	q_n:=\left(\frac{\hbar^4\epsilon}{2m^2}\right)(n+\frac{1}{2})^2
	(11 n^2+9n+4)\,.
	\label{g5}
	\ee
It is not difficult to show that Eq.~(\ref{g4}) has a single positive solution
[This is true for any positive $p_n$ and $q_n$.] given by \cite{galois}
	\be
	{\cal E}=\left(\frac{q_n}{2}\right)^{1/3}\left(1+\sqrt{1-r_n}\right)^{1/3}
	+\left(\frac{p_n}{3}\right)\left(\frac{2}{q_n}\right)^{1/3}
	\left(1+\sqrt{1-r_n}\right)^{-1/3}\;,
	\label{g6}
	\ee
where
	\be
	r_n:=\frac{4p_n^3}{27q_n^2}=
	\left(\frac{8n+4}{11n^2+9n+4}\right)^2r_0\;,
	~~~~ r_0:=\frac{m k^3}{108\hbar^2\epsilon^2}\;.
	\label{g6.1}
	\ee

The right-hand side of Eq.~(\ref{g6}) is manifestly real and positive for 
$r_n\leq 1$. It is not difficult to check that it is also real and positive for 
$r_n>1$. In fact, we can express ${\cal E}$ in the form
	\be
	{\cal E}=2 \sqrt{\frac{p_n}{3}} \cos\left(\frac{\phi_n}{3}\right)
	=\sqrt{\frac{ k}{3m}}(2n+1)
	\cos\left(\frac{\phi_n}{3}\right)\;,
	\label{g7}
	\ee
where
	\be
	\phi_n:=\tan^{-1}(\sqrt{r_n-1})\;.
	\label{g8}
	\ee
Note that for $r_n<1$, $\phi_n$ is imaginary, but $\cos(\phi_n/3)$ is still real 
and positive.

Having fixed the parameter ${\cal E}$, we can determine the energy eigenvalues 
$E_n$ using Eqs.~(\ref{f7}) and (\ref{g3.1}). We first use 
Eqs.(\ref{alpha=}) and~(\ref{g7}) to compute
	\be
	\alpha_n=\left(\frac{2}{\hbar}\right)\sqrt{\frac{m k}{3}}
	\cos\left(\frac{\phi_n}{3}\right)\;.
	\label{g9}
	\ee
Then substituting this equation in Eq.~(\ref{g3.1}) and using Eq.~(\ref{f7}), 
we find
	\be
	E_n= \left(\frac{\hbar}{24}\right)\sqrt{\frac{3 k}{m}}(2n+1)
	\left[7+3\tan^2(\frac{\phi_n}{3})\right]\cos(\frac{\phi_n}{3})+
	\left(\frac{9\hbar^2\epsilon}{16m k}\right)(2n^2+2n+1)
	\left[1+\tan^2(\frac{\phi_n}{3})\right]\;.
	\label{g10}
	\ee
In particular, the ground state energy is given by
	\be E_0= \left(\frac{\hbar}{24}\right)\sqrt{\frac{3 k}{m}}
	\left[7+3\tan^2(\frac{\phi_0}{3})\right]\cos(\frac{\phi_0}{3})+
	\left(\frac{9\hbar^2\epsilon}{16m k}\right)
	\left[1+\tan^2(\frac{\phi_0}{3})\right]\;.
	\label{g10.1}
	\ee

In Table~1, we list the numerical values obtained using Eq.~(\ref{g10})
for the first 10 energy levels of a quartic anharmonic oscillator with $m=1/2$,
$k=2$, $\epsilon=1/10$ in units where $\hbar=1$. This table also includes the
accurate numerical values of Ref.~\cite{bacus}, the values obtained
using the conventional Sturmian approximation and the zero and first order 
perturbation theory.[The zero and first order perturbation theory yield 
	\be
	E_n^{(0)}=\hbar( k/m)^{1/2}(n+1/2)\;,~~~~
	E_n^{(1)}=E_n^{(0)}+3\hbar^2\epsilon^2(2n^2+2n+1)/(4m k)\,,
	\label{pert-2-4}
	\ee
respectively.] The relative difference between the results of the variational 
Sturmian approximation of order zero with the highly accurate numerical results 
($E_n^\#$) of Ref.~\cite{bacus}, i.e., the quantity $|E_n-E_n^\#|/E^\#_n$, 
varies between $3.38\times 10^{-4}$ and $2.77\times 10^{-3}$. For the 
ground state, this number is $1.53\times 10^{-3}$. Even for the lowest lying 
energy levels where perturbation theory yields reliable results, the zero order 
variational Sturmian approximation produces more accurate values than both 
the zero and first order perturbation theory. As seen from Table~1, the
variational Sturmian approximation is better than the conventional Sturmian 
approximation. 

In the remainder of this section we present the results obtained using the
first and second order variational Sturmian approximation. The numerical 
results are respectively presented in Tables~2 and~3.

As we explained in Section~IV, in the variational Sturmian approximation of
order one one chooses an indexing set ${\cal S}_2$ consisting of two 
Sturmians to be included in the expansion of the eigenvector $|E\kt$.
One then solves the corresponding secular equation~(\ref{e12}), expresses the solutions 
$E_\pm$ in terms of the parameter ${\cal E}$, and finds the minima of 
$E_\pm({\cal E})$. In general, $E_\pm$ are complicated functions of 
${\cal E}$. However, it turns out that for all the cases that we considered
$E_\pm$ has a unique minimum corresponding to a positive value of 
${\cal E}$. 

In order to choose the indexing set ${\cal S}_2$, we first note that the 
Sturmian functions $\br x|\phi_n\kt$ with even (respectively odd) $n$ are even 
(respectively odd) functions of $x$. We expect the energy eigenfunctions
of the anharmonic oscillator~(\ref{g1}) to have the same parity structure as the
Sturmian functions. This, in particular, suggests that in the calculation of $E_0$ 
we should take  ${\cal S}_2=\{0,2\}$. 

For a quartic anharmonic oscillator with $m=1/2$, $ k=2$, 
$\epsilon=1/10$, the first order variational Sturmian approximation 
corresponding to ${\cal S}_2=\{0,2\}$ yields $E_0=E_-=1.06614$ and 
$E_2=E_+=5.76117$. The value obtained for $E_0$ differs from the accurate 
numerical value by one part in $10^4$. It is one order of magnitude better than 
the value obtained using the zero order variational Sturmian approximation. 
The value for $E_2$ is however less accurate. One may argue that the choice 
made for ${\cal S}_2$ is appropriate only for the ground state. In order to 
compute $E_2$ using the first order variational Sturmian approximation, one 
may alternatively choose ${\cal S}_2=\{2,4\}$. This choice yields 
$E_2=E_-=5.74558$ and $E_4=E_+=9.6637$. Again this value for $E_2$ is 
an order of magnitude better than the value obtained using the zero order 
variational Sturmian approximation, whereas the value for $E_4$ is less 
accurate.  One can also try ${\cal S}_2=\{0,4\}$. As expected, this choice 
yields a less accurate value than the choices ${\cal S}_2=\{0,2\}$ and 
${\cal S}_2=\{2,4\}$ for both $E_0$ and $E_4$. 

For the calculation of the first excited state we choose ${\cal S}_2=\{1,3\}$. 
Then we find $E_1=E_-=3.30922$ and $E_3=E_+=8.37284$. Once again the 
first order variational Sturmian approximation of order one with the choice 
${\cal S}_2=\{1,3\}$ yields a more accurate result for $E_1$ and a less 
accurate result for $E_3$.

In general, in the calculation of the energy levels $E_n$ with $n\geq 2$, there 
are two alternative choices for the indexing set ${\cal S}_2$. In view of the 
parity properties of the eigenvectors, these are $\{n,n+2\}$ and $\{n-2,n\}$. 
The fact that there is no physical reason to distinguish between these two 
choices suggests that for these levels one should consider the second order 
variational Sturmian approximation with the choice 
${\cal S}_3=\{n-2,n,n+2\}$. 

Table~3 includes the results of the second order variational Sturmian 
approximation corresponding to the indexing set ${\cal S}_3=\{0,2,4\}$.  
This approximation yields more accurate values for $E_0$ than the zero and 
first order variational Sturmian approximations. However, contrary to our
expectation the value obtained for $E_2$ is less accurate than the one given by 
the zero order approximation and the first order approximation with 
${\cal S}_2=\{2,4\}$.  

\subsection*{The Quartic Potential}

Consider the quartic potential
	\be
	V(x)=\epsilon\,x^4\;.
	\label{h1}
	\ee
We can easily obtain  the energy levels of this potential using the zero order
variational Sturmian approximation by simply setting $ k=0$ in our 
formulas for the quartic anharmonic oscillator. Substituting $ k=0$
in (\ref{g5}), we can write Eq.~(\ref{g4}) in the form
	\be
	{\cal E}=q_n^{1/3}=\hbar\left(\frac{\hbar\epsilon}{2m^2}\right)^{1/3}
	\left[(n+\frac{1}{2})^2(11n^2+9n+4)\right]^{1/3}\;.
	\label{h2}
	\ee
In view of Eqs.~ (\ref{alpha=}, (\ref{f7}), (\ref{g3.1}), (\ref{h2}), and 
$ k=0$, we have
	\bea
	\alpha_n&=&\hbar^{-1}\left(m\hbar\epsilon\right)^{1/3}
	\left(\frac{11n^2+9n+4}{2n+1}\right)^{1/3}\;,
	\label{h3}\\
	E_n&=&\left(\frac{8n+4}{11n^2+9n+4}\right)^{2/3}
			\left(\frac{17}{7}n^2+\frac{15}{7}n+1\right)E_0\;,
	\label{h4}\\
	E_0&:=&
	\frac{7\hbar}{8}\left(\frac{\hbar\epsilon}{2m^2}\right)^{1/3}\;.
	\label{h5}
	\eea

In Table~4, we present the values obtained using Eq.~(\ref{h4}) for the 
energy levels of a quartic potential with $m=1/2$ and $\epsilon=1$ in 
units where $\hbar=1$. This table also includes accurate numerical results 
$E_n^\#$ and the results of the zero and first order WKB approximation 
given in Refs.~\cite{bender,voros}. The relative difference 
$|E_n-E_n^\#|/E_n^\#$ is about $0.04$ for the ground state
and ranges between $6.4\times 10^{-4}$ and $8.7\times 10^{-3}$ for the energy 
levels $E_2,E_4,E_6,E_8,E_{10}$ and $E_{16}$. 

Table~5 includes the results of the first and second order variational 
Sturmian approximation for $E_0, E_2,$ and $E_4$.

\subsection*{The Gaussian Potential}

Consider the Gaussian potential
	\be
	V(x)=-\lambda\,e^{-\epsilon x^2/2}\;,.
	\label{i1}
	\ee
In order to apply the results of section~IV to this potential, we write 
$V(x)=\tilde V(x)-\lambda$ where
	\be
	\tilde V(x)=\lambda(1-e^{-\epsilon x^2/2})\;.
	\label{i2}
	\ee
Then as $\epsilon$ tends to zero, $\tilde V(x)$ approaches to the harmonic
oscillator potential (\ref{f1}) with $k=\lambda\epsilon$.

Clearly, the energy eigenvalues associated with $V$ and $\tilde V$ are related by
	\be
	E_n=\tilde E_n-\lambda\;.
	\label{i3}
	\ee
In the following we use the zero order variational Sturmian approximation to
obtain the ground state energy of the potential $\tilde V$. The excited energy
levels can be obtained similarly.

We first note that for the ground state $n=0$, and Eq.~(\ref{f14}) for the potential 
$\tilde V$ takes the form
	\bea
	{\cal E}&=&\sqrt{2}~_0\br 0|\tilde V|2\kt_0 =
	\sqrt{2}~_0\br 0|V|2\kt_0=
	-\sqrt{2}\lambda  ~_0\br 0|e^{-\epsilon x^2/2}|2\kt_0\nn\\
	&=&-\sqrt{2}\lambda\int_{-\infty}^\infty
	~_0\br0|x\kt e^{-\epsilon x^2/2}\br x|2\kt_0\,dx\;.
	\label{i4}
	\eea
We can evaluate the right-hand side of (\ref{i4}) using the well-known 
expression for the eigenfunctions of the harmonic oscillator, namely 
(\ref{f1.4}) and
	\be
	\br x|2\kt_0=\left(\frac{\alpha_0}{4\pi}\right)^{1/4}(2\alpha_0x^2-1)	
	e^{-\alpha_0x^2/2}\;.
	\label{i5}
	\ee
Substituting Eqs.~(\ref{f1.4}) and (\ref{i5}) in Eq.~(\ref{i4}) and performing 
the necessary calculations, we find
	\be
	{\cal E}({\cal E}+p)^3=\lambda^2 p^2\;,
	\label{i6}
	\ee
where 
	\be
	p:=\frac{\hbar^2\epsilon}{4m}\;.
	\label{i7}
	\ee

Introducing 
	\be
	\eta:=1+\frac{\cal E}{p}\,,
	\label{i7.1}
	\ee
we can write Eq.~(\ref{i6}) in the form
	\be
	f(\eta):=\eta^4-\eta^3-r=0\;,
	\label{i8}
	\ee
where 
	\be
	r:=\frac{\lambda^2}{p^2}=
	\frac{16\lambda^2 m^2}{\hbar^4\epsilon^2}\;.
	\label{i9}
	\ee
It is not difficult to show that for all $r>0$, $f(\eta)$ has a single minimum at 
$\eta=3/4$. The minimum is $f(3/4)=-(27/256+r)<0$. Furthermore, 
$f(0)=f(1)=-r<0$ and $\lim_{\eta\to\infty} f(\eta)=\infty$. Therefore, $f(\eta)$ 
has a single positive root that is greater than $1$. This root is given by
	\be
	\eta_\star=\frac{1}{4} \left(1+2\xi+\sqrt{3-4\xi^2+\xi^{-1}}\right)\;,
	\label{i10}
	\ee
where
	\bea
	\xi&:=&\frac{1}{2}\sqrt{1-a+b},~~~~
	a:= \frac{3}{2}\,\zeta(1+\sqrt{1+\zeta^3})^{1/3} ,\nn\\
	b&:=&\frac{3}{2}\,\zeta(-1+\sqrt{1+\zeta^3})^{1/3} ,~~~~
	\zeta:=\frac{4}{3}\,(4r)^{1/3}\;.\nn
	\eea
In view of Eq.~(\ref{i7.1}) and the fact that $\eta_0>1$, Eq.~(\ref{i6}) has
a single positive solution, namely
	\be
	{\cal E}=p(\eta_\star-1).
	\label{i11}
	\ee

Having obtained the parameter ${\cal E}$, we next compute
	\be
	W=~_0\br 0|\tilde V|0\kt_0= 
	\lambda\left(1-\sqrt{1-\eta_\star^{-1}}\right) \;.
	\label{i12}
	\ee
Here we have made use of Eqs.~(\ref{f1.4}), (\ref{alpha=}), (\ref{i2}),
(\ref{i7}), and (\ref{i11}). Substituting this equation and Eq.~(\ref{i11}) in
Eq.~(\ref{f7}) and using Eqs.~(\ref{i3}) and (\ref{i9}), we find the ground 
state energy of the Gaussian potential (\ref{i1}) to be
	\be
	E_0=\tilde E_0-\lambda=
	-\lambda\left[\sqrt{1-\eta_\star^{-1}}+
	\frac{1-\eta_\star}{2\sqrt{r}}\right]\;.
	\label{i13}
	\ee

In order to reveal the asymptotic behavior of $E_0$, we investigate the power 
series expansion of the right-hand side of Eq.~(\ref{i13}).
	\begin{itemize}
	\item[~] For $r\gg 1$, i.e., $\frac{\epsilon}{\lambda}\to 0$,
	\be
	E_0=-\lambda[1-r^{-1/4}+\frac{3}{8}\,r^{-1/2}-\frac{1}{32}\,r^{-3/4}
	-\frac{1}{128}\,r^{-1}+{\cal O}(r^{-5/4})]\;.
	\label{i14}
	\ee
	\item[~] For $r\ll 1$, i.e., $\frac{\epsilon}{\lambda}\to \infty$,
	\be
	E_0=-(\frac{\lambda\sqrt{r}}{2})
	[1-r+3r^2-13r^3+68 r^4+{\cal O}(r^5)]\;.
	\label{i15}
	\ee
	\end{itemize}
Therefore, for fixed $\lambda$,
	\bea
	\lim_{\epsilon\to 0^+} E_0&=&-\lambda\;,
	\label{i16.1}\\
	\lim_{\epsilon\to\infty} E_0&=& 
	\lim_{\epsilon\to\infty}\left[-\frac{4m^2
	\lambda^2}{\hbar^2\epsilon}\right]=0^-\;,
	\label{i16.2}
	\eea
and for fixed $\epsilon$,
	\bea
	\lim_{\lambda\to 0^+} E_0&=&\lim_{\lambda\to 0}\left[-\frac{4m^2
	\lambda^2}{\hbar^2\epsilon}\right]=0^-\;,
	\label{i17.1}\\
	\lim_{\lambda\to\infty} E_0&=&-\infty\;.
	\label{i17.2}
	\eea
Clearly, the asymptotic behavior of $E_0$, as given by Eqs.~(\ref{i16.1}) --
(\ref{i17.2}), agrees with the qualitative analysis of the eigenvalue problem for
the Gaussian potential.

It is not difficult to see that in the limit $\epsilon\to 0$ perturbation theory 
provides reliable results. Writing the Gaussian potential~(\ref{i1}) in the form
	\be
	V=V_0+\delta V\;,
	\label{i18}
	\ee
with $V_0$ given by Eq.~(\ref{sho}) and performing the standard calculations
\cite{perturbation}, we find that the zero and first order perturbation theory yield
respectively	
	\bea
	E_0^{(0)}&=&-\lambda(1-r^{-1/4})\;,
	\label{i19.1}\\
	E_0^{(1)}&=&-\lambda(1+r^{-1/4})^{-1/2}=
	-\lambda[1-\frac{1}{2}\,r^{-1/4}+\frac{3}{8}\, 
	r^{-1/2}-\frac{5}{16}\,r^{-3/4}+{\cal O}(r^{-1})]\;.
	\label{i19.2}
	\eea
Here $E_0^{(\ell)}$ is the ground state energy for the Gaussian 
potential~(\ref{i1}) obtained using the $\ell$-th order perturbation theory. 

Comparing Eq.~(\ref{i14}) with Eqs.~(\ref{i19.1}) and (\ref{i19.2}), one finds 
that in the pertubative region where $r\gg 1$, the variational Sturmian 
approximation of order zero agrees with the results of the perturbation 
theory. In fact, since by construction $E_0$ is the expectation value of the energy of 
the Sturmian $|\phi_0\kt$, the fact that $E_0<E_0^{(0)}$ shows that even in 
the pertubative region the variational Sturmian approximation of order zero is a 
better approximation than the zero order perturbation theory. By the same reasoning,
because $E_0>E_0^{(1)}$, the first order perturbation theory yields a better result.
Note however that the wave function obtained in the first order perturbation theory
is an infinite sum whereas the wave function in the zero order Sturmian approximation is
given explicitly.

Another interesting limit is the delta function limit of the potential $V$ where 
$\lambda=a\sqrt{\epsilon/(2\pi)}$, $\epsilon\to\infty$, and $V(x)\to -a\delta(x)$. 
Here $a$ is a fixed coupling constant. In this limit $r\to 0$ and the ground state
energy is given by Eq.~(\ref{i15}) according to
	\be
	E_0=-\frac{m a^2}{\pi\hbar^2}\;.
	\label{i20}
	\ee
This result has the same order of magnitude as the exact result: 
	\be
	E_0=-\frac{m a^2}{2\hbar^2}\;.
	\label{i21}
	\ee

\section*{VI.~Discussion and Conclusion}

We have outlined a variationally improved Sturmian approximation and applied our
results to the harmonic oscillator Sturmians. For these Sturmians we could solve
the associated variational problem in the zero order Sturmian approximation exactly.
We have used our variational Sturmian approximation in the calculation of the energy
levels of various potentials. We have shown that using a few harmonic oscillator 
Sturmians, one obtains quite reliable results. In general, the variational Sturmian
approximation is a better approximation than the conventional Sturmian
approximation.

Because the harmonic oscillator potential is a confining potential, we expect that 
the method is more suitable for the confining potentials such as the quartic 
anharmonic oscillator and the quartic potential. We can base this argument on a
more quantitative reasoning by addressing the problem of classifying the potentials for 
which the Sturmian approximation is exact. It is not difficult to show that these 
potentials satisfy
	\be
	V(\vec x)=E-{\cal E}+\left(
	\frac{ \sum_{\nu=0}^N\sum_\alpha
	C_\nu^\alpha\beta_\nu\phi_{\nu,\alpha}(\vec x) }{
	\sum_{\nu=0}^N
	 C_\nu^\alpha\phi_{\nu,\alpha}(\vec x)}\right)V_0(\vec x)\;,
	\label{j1}
	\ee
where $E$, ${\cal E}$ and $C_\nu^\alpha$ are constants and 
$\phi_{\nu,\alpha}(\vec x):=\br \vec x|\phi_\nu,\alpha\kt$. Eq.~(\ref{j1})  follows 
from Eqs.~(\ref{xe1}), (\ref{xe2}), (\ref{e3}), and (\ref{e6}). 

For example, the potentials for which the first order harmonic oscillator Sturmian
approximation with ${\cal S}_2=\{0,2\}$ yields an exact eigenfunction are of the form
	\be
	V(x)=E-\frac{\hbar^2\alpha_0}{2m}+
	\left(\frac{\hbar^2\alpha_0^2}{2m}\right)
	\left[\frac{e^{-2\alpha_0 x^2/5}+
	(\frac{\zeta}{5})(2\alpha_0\,x^2-5)}{
	e^{-2\alpha_0 x^2/5}+5 \zeta(2\alpha_0\,x^2-5)}\right]x^2\,,
	\label{j2}
	\ee
where $\alpha_0$ is a real parameter with the dimension of (length$)^{-2}$ and 
$\zeta$ is a dimensionless real parameter. As seen from Eq.~(\ref{j2}), these
potentials tend to the harmonic oscillator potential for $|x|\to\infty$. In particular,
as $|x|\to\infty$, $V\to\infty$. This asymptotic behaviour is also valid for higher order
harmonic oscillator Sturmian approximations. This observation shows that the 
harmonic oscillator Sturmian approximation is more reliable for confining 
potentials. 

We conclude this paper with a couple of remarks. 
	\begin{itemize}
	\item[1.] The variational principle used in the variational Sturmian 
approximation leads to an algebraic (nondifferential) equation for the 
parameter ${\cal E}$. The acceptable solutions for this equation are those 
which are real and positive. The fact that for all the cases we consider there
is a unique real positive solution corresponding to each eigenvalue $E_n$ is quite
remarkable. This observation may be viewed as a consistency check for the
Sturmian approximation. 
	\item[2.] In our selection of the Sturmians in the first and higher order
Sturmian approximation, we used the information about the parity
properties of the Sturmians and the energy eigenfunctions. For example we
ruled out the first order variational Sturmian approximation with ${\cal S}_2=
\{0,1\}$. If we perform the necessary calculations, we find that for this
choice the functions $t$ and $w$ vanish identically and the matrices $T$ and
$S$ are diagonal. Therefore, the secular equation (\ref{e12}) yields the same
results as the zero order Sturmian approximation. This can also be seen from
the results of Ref.~\cite{antonsen}. 
	\end{itemize}

\newpage

\begin{table}
	\begin{center}
	\begin{tabular}{ | r | r | r | r | r | r | r | r | r | r | }
	\hline
	$n$ &
	$E_n^\#$&
	$E_n$ &	
	$\frac{|E_n-E_n^\#|}{E^\#_n}$ &
	$E_n^{\rm CSA}$ & 
	$\frac{|E^{CSA}_n-E_n^\#|}{E^\#_n}$ &
	$E_n^{(0)}$ & 
	$\frac{|E^{(0)}_n-E_n^\#|}{E^\#_n}$ &
	$E_n^{(1)}$ & 
	$\frac{|E^{(1)}_n-E_n^\#|}{E^\#_n}$ \\
	\hline
	0 & 1.065286 & 
	1.06692 & $1.5\times 10^{-3}$ &
	1.07500 & $9.1\times 10^{-3}$ &
	1.000 & 0.061 &
	1.075 & $9.1\times 10^{-3}$ \\
	\hline
	1 & 3.306872 & 
	3.31182 & $1.5\times 10^{-3}$ &
	3.37500 & 0.021 &
	3.000 & 0.032 &
	3.450 & 0.043 \\
	\hline
	2 & 5.747959 & 
	5.75052 & $4.5\times 10^{-4}$ &
	5.97500 & 0.040 &
	5.000 & 0.13 &
	5.975 & 0.039 \\
	\hline
	3 & 8.352678 & 
	8.34985 & $3.4\times 10^{-4}$ &
	8.87500 & 0.063 &
	7.000 & 0.16 &
	8.875 & 0.063 \\
	\hline
	4 & 11.09860 & 
	11.0881 & $9.5\times 10^{-4}$ &
	12.0750 & 0.088 &
	9.000 & 0.19 &
	12.08 & 0.088 \\
	\hline
	5 & 13.96993 & 
	13.9499 & $1.4\times 10^{-3}$ &
	- & - &
	11.00 & 0.21&
	15.58 & 0.11 \\
	\hline
	6 & 16.95479 & 
	16.9235 & $1.8\times 10^{-3}$ &
	- & - &
	13.00 & 0.23 &
	19.38 & 0.14 \\
	\hline
	7 & 20.04386 &
	19.9998 & $2.2\times 10^{-3}$ &
	- & - &
	15.00 & 0.25 &
	23.48 & 0.17 \\
	\hline
	8 & 23.22955 & 
	23.1715 & $2.5\times 10^{-3}$ &
	- & - &
	17.00 & 0.27 &
	27.88 & 0.20 \\
	\hline
	9 & 26.50555 & 
	26.4322 &$2.8\times 10^{-3}$ &
	- & - &
	19.00 & 0.28 &
	32.58 & 0.23 \\
	\hline
	\end{tabular}
	\caption{First 10 energy levels of the Hamiltonian
	$H=p^2+x^2+\frac{x^4}{10}$ in units where $\hbar=1$. 
	$E_n^{\#}$ are the highly accurate numerical values of 
	Ref.~\cite{bacus}. $E_n$ are the values obtained using the zero order
	variational Sturmian approximation. $E_n^{\rm CSA}$ are the values
	obtained by the zero order conventional Sturmian approximation in
	Ref.~\cite{antonsen}. $E_n^{(0)}$ and $E_n^{(1)}$ are
	the energy eigenvalues obtained using the zero and first order
	perturbation theory, respectively.}
	\end{center}
\end{table}


{\footnotesize
\begin{table}
\begin{center}
	\begin{tabular}{ | r | r | r | r | r | r | r | r | r | r | r | }
	\hline
	${\cal S}_2$ &
	$E_0$ &	$\delta E_0$ &
	$E_1$ &	$\delta E_1$ &
	$E_2$ &	$\delta E_2$ &
	$E_3$ &	$\delta E_3$ &
	$E_4$ &	$\delta E_4$ \\
	\hline
	\{0,2\} &
	1.06614 & $8.0\times 10^{-4}$&
	- & - &
	5.76117 & $2.3\times 10^{-3}$&
	- & - &
	- & - \\
	\hline
	\{1,3\} &
	- & - &
	3.30922 & $7.1\times 10^{-4}$&
	- & - &
	8.37284 & $2.4\times 10^{-3}$&
	- & - \\
	\hline
	\{2,4\} &
	- & - &
	- & - &
	5.74558 & $4.1\times 10^{-4}$&
	- & - &
	9.66370& 0.13 \\
	\hline
	\{0,4\} &
	1.06620 & $8.6\times 10^{-4}$&
	- & - &
	- & - &
	- & - &
	9.64502& 0.13 \\
	\hline
	\end{tabular}
	\caption{Energy levels of the Hamiltonian
	$H=p^2+x^2+x^4/10$ obtained using the first order variational
	Sturmian approximation. $\delta E_n$ stands for $|E_n-E_n^\#|/
	E^\#_n$.}
	\end{center}
\end{table}
}


\begin{table}
\begin{center}
	\begin{tabular}{ | r | r | r | r | r |}
	\hline
	$n$ & $E_n^\#$& $E_n$ & $\frac{|E_n-E_n^\#|}{E_n^\#}$\\
	\hline
	0 & 1.065286 &
	1.06613 & $7.9\times 10^{-4}$\\
	\hline
	2 & 5.75052 &
	5.75275 & $8.3\times 10^{-4}$\\
	\hline
	4 & 11.09860 &
	9.68483 & $0.127$\\
	\hline
	\end{tabular}
	\caption{Energy levels of the Hamiltonian
	$H=p^2+x^2+x^4/10$ obtained using the second order variational
	Sturmian approximation with the choice $\{0,2,4\}$ for the indexing set 
	${\cal S}_3$. $E_n^\#$ are the highly accurate numerical 
	values of Ref.~\cite{bacus}. }
	\end{center}
\end{table}


\begin{table}
	\begin{center}
	\begin{tabular}{ | r | r | r | r | r | r | r | r | }
	\hline
	$n$ &
	$E_n^\#$ &
	$E_n$ &	
	$\frac{|E_n-E_n^\#|}{E^\#_n}$ &
	$E_n^{(0)}$ & 
	$\frac{|E^{\rm WKB(0)}_n-E_n^\#|}{E^\#_n}$ &
	$E_n^{(1)}$ & 
	$\frac{|E^{\rm WKB(1)}_n-E_n^\#|}{E^\#_n}$ \\
	\hline
	0 & 1.060362 & 
	1.10243 & 0.040&
	0.87 & 0.17&
	0.98 &  0.076\\
	\hline
	1 & - & 
	3.86929 & - &
	- & - &
	- &  - \\
	\hline
	2 & 7.455697 & 
	7.46048 & $6.4\times 10^{-4}$ &
	7.4140 & $5.6\times 10^{-3}$&
	7.4558 & $1.4\times 10^{-5}$\\
	\hline
	3 & - & 
	11.6007 & - &
	- & - &
	- & - \\
	\hline
	4 & 16.261826 & 
	16.1691 & $5.7\times 10^{-3}$ &
	16.233615 & $1.7\times 10^{-3}$ &
	16.261937 & $6.8\times 10^{-6}$  \\
	\hline
	6 & 26.528471 & 
	26.3349 & $7.3\times 10^{-3}$ &
	26.506336 & $8.3\times 10^{-4}$ &
	26.528513 & $1.9\times 10^{-5}$ \\
	\hline
	8 & 37.923001 & 
	37.6218 & $7.9\times 10^{-3}$ &
	37.904472 & $4.9\times 10^{-4}$ &
	37.923021 & $5.3\times 10^{-7}$  \\
	\hline
	10 & 50.256255 & 
	49.8404 & $8.3\times 10^{-3}$ &
	50.240152 & $3.1 \times 10^{-4}$  &
	50.256266 & $2.2\times 10^{-7}$ \\
	\hline
	16 & 91.79806 & 
	91.0012 & $8.7\times 10^{-3}$ &
	- & - &
	- & - \\
	\hline
	\end{tabular}
	\caption{Energy levels of the Hamiltonian
	$H=p^2+x^4$ in units where $\hbar=1$. 
	$E_n^\#$ are the highly accurate numerical values of 
	Refs.~\cite{bender,voros}. $E_n$ are the values obtained using the zero 
	order variational Sturmian approximation. $E_n^{\rm WKB(0)}$ and 
	$E_n^{\rm WKB(1)}$ are the values obtained using the zero and first 
	order WKB approximation \cite{bender}, respectively.}
	\end{center}
\end{table}


\begin{table}
\begin{center}
	\begin{tabular}{ | r | r | r | r | r | r | r | r | }
	\hline
	N & ${\cal S}_N$ &
	$E_0$ &	$\frac{|E_0-E_0^\#|}{E^\#_0}$ &
	$E_2$ &	$\frac{|E_2-E_2^\#|}{E^\#_2}$ &
	$E_4$ &	$\frac{|E_4-E_4^\#|}{E^\#_4}$  \\
	\hline
	2 & \{0,2\} &
	1.08110 & $0.0196$&
	7.60884 & $0.0205$&
	- & - \\
	\hline
	2 & \{2,4\} &
	- & - &
	7.42669 & $3.89\times 10^{-3}$&
	16.4461 & $0.0113$\\
	\hline
	2 & \{0,4\} &
	1.08166& 0.0200 &
	- & - &
	16.4114 & $9.12\times 10^{-3}$\\
	\hline
	3 & \{0,2,4\} &
	1.08010 & $0.0195$&
	7.56528 & $0.0147$ &
	16.5670 & $0.0188$ \\
	\hline
	\end{tabular}
	\caption{Energy levels of the Hamiltonian
	$H=p^2+x^4$ obtained using the first and second order variational
	Sturmian approximation. $N$ is the order of the approximation. 
	$E_n^\#$ are the accurate numerical results reported in 
	Ref.~\cite{bender}.}
	\end{center}
\end{table}

\end{document}